\documentclass[conference]{IEEEtran}
\usepackage{epstopdf}
\usepackage{epsfig}
\usepackage{graphicx,dblfloatfix}
\usepackage{tabularx}
\usepackage{amsmath}
\usepackage{amssymb}
\usepackage{relsize} 
\usepackage{wasysym}
\usepackage{graphicx}
\usepackage{tikz}
\usetikzlibrary{calc,positioning,shapes.geometric}
\usetikzlibrary{trees,shapes}
\usepackage{subfig}
\usepackage{lscape}
\usepackage{pdflscape}
\usepackage{xcolor,colortbl}
\usepackage{url}
\usepackage{balance}
\usepackage[square,sort,comma,numbers]{natbib}
\begin{document}

\title{Using Source Code Metrics and Ensemble Methods for Fault Proneness Prediction}

\author{\IEEEauthorblockN{Lov Kumar}
\IEEEauthorblockA{NIT Rourkela, India\\
lovkumar505@gmail.com}
\and
\IEEEauthorblockN{Santanu Rath}
\IEEEauthorblockA{NIT Rourkela, India\\
skrath@nitrkl.ac.in}
\and
\IEEEauthorblockN{Ashish Sureka}
\IEEEauthorblockA{ABB Corporate Research, India\\
ashish.sureka@in.abb.com}
} 
\maketitle
\begin{abstract}
Software fault prediction model are employed to optimize testing resource allocation by identifying fault-prone classes before testing phases. Several researchers' have validated the use of different classification techniques to develop predictive models for fault prediction. The performance of the statistical models are proven to be influenced by the training and testing dataset. Ensemble method learning algorithms have been widely used because it combines the capabilities of its constituent models towards a dataset to come up with a potentially higher performance as compared to individual models (improves generalizability). In the study presented in this paper, three different ensemble methods have been applied to develop a model for predicting fault proneness. The efficacy and usefulness of a fault prediction model also depends on the source code metrics which are considered as the input for the model. 

In this paper, we propose a framework to validate the source code metrics and select the right set of metrics with the objective to improve the performance of the fault prediction model. The fault prediction models are then validated using a cost evaluation framework. We conduct a series of experiments on 45 open source project dataset. Key conclusions from our experiments are: (1) Majority Voting Ensemble (MVE) methods outperformed other methods; (2) selected set of source code metrics using the suggested source code metrics using validation framework as the input achieves better results compared to all other metrics; (3) fault prediction method is effective for software projects with a percentage of faulty classes lower than the threshold value (low - 54.82\%, medium - 41.04\%, high - 28.10\%)
\end{abstract}
\IEEEpeerreviewmaketitle
\section{Introduction}
Early identification of source code regions, classes or modules where faults are likely to occur can help in optimizing and guiding testing efforts resulting in improvement of software quality. Fault prediction is an important and challenging problem in the area of software engineering and hence attracted the attention of several researchers. Many fault prediction techniques have been proposed and their performance have been evaluated on different dataset. Hall et al. \cite{hall2012} and Catal et al. \cite{catal2009} conduct a systematic literature review in the area of fault prediction. Hall et al. analyze 208 fault prediction studies and conclude that the methodology used to build models seems to be influential to predictive performance. One of the conclusions of their study is that more studies are needed that use a reliable methodology and which report their context, methodology, and performance comprehensively \cite{hall2012}. Arisholm et al. conduct a comprehensive examination of methods to develop and evaluate the performance of fault prediction models \cite{arisholm2010}. Their study is performed in an industrial setting in the context of large Java legacy system development project \cite{arisholm2010}.

We believe that there are several research gaps and scope for more studied in the area of fault prediction. One of the research gaps is in modeling technique.  The application of homogeneous and heterogeneous ensemble methods is relative unexplored. One of the novel contributions of our work in context to existing work is the application of various base learners such as logistic regression analysis, artificial neural network and radial basis function neural network as constituents of ensemble methods. Another research contribution of the work presented in this paper is an in-depth study of the generalizability of fault prediction model as we conduct our experiments on dataset belonging to 45 open source projects. We propose a cost evaluation model which takes into account the economics of software quality assurance \cite{a35}. The application of estimated fault removal cost, testing cost and the normalized fault removal cost in the fault prediction framework based on ensemble methods is a novel and unique contribution of our work.\\

\noindent\textbf{Note:} The research presented in this paper is an extended version of the paper published by the same authors in COMPSAC 2017 \cite{compsac2017}. The COMPSAC 2017 paper is a short paper (6 pages) and hence does not contain all the details due to the page limit. 

\section{Background}
\label{ResBackground}
In this section, we define the dependent and independent variables of the models and provide the experimental dataset description. 
\subsection{Dependent and Independent Variables}
In our study, the objective is to develop a fault prediction model using source code metrics as input variables or predictors. Hence, the class or category "bugs" is a dependent variable and various sets of metrics based on source code are independent
variables. The  independent and dependent variables for the fault prediction model is shown in Table \ref{ana}. We compute the source code metrics listed in Table \ref{ana} for the purpose of building a predictive model for estimating fault-proneness of the software system. We compute the source code metrics using CKJM extended tool\footnote{\url{http://gromit.iiar.pwr.wroc.pl/p_inf/ckjm/}} which is an extended version of the CKJM tool for calculating Chidamber and Kemerer Java Metrics and several other metrics such as WMC (Weighted methods per class), LCOM (Lack of cohesion in methods), CBO (Coupling between object classes), MFA (Measure of Functional Abstraction), CAM (Cohesion Among Methods of Class) and CC (McCabe cyclomatic complexity) \cite{chidamber1991}. We apply Chidamber and Kemerer Java Metrics in our study as these metrics have been widely and successfully used for predicting change-proneness of object oriented system (a different problem than fault prediction but in the domain of using source code metrics for predicting software quality and maintainability) and has tool support for the purpose of collecting the relevant metric data from source code \cite{kumar2016quasoc}\cite{kumar2017isec}\cite{kumar2017malt}.
\begin{table}[h]
	\centering
	\caption{Predictors and Target Class}
	
	\renewcommand{\arraystretch}{1.1}
	\resizebox{8.5cm}{!}{
	\begin{tabular}{|p{1.70cm}|p{7cm}|}
		\hline
		\textbf{Dependent Variable} & \textbf{Independent Variable}  \\
		\hline
		
		Fault & DIT, WMC, RFC, CBO, LCOM,  NOC, Ce, Ca, LCOM3, NPM,  DAM, MOA, LOC, CAM, MFA, CBM, AVG-CC, MAX-CC, AMC, IC
		\\ \hline
		Fault & Reduced feature attributes using proposed framework \\ \hline
	\end{tabular}}
	
	\label{ana}
\end{table}

\subsection{Experimental Dataset}
Based on our analysis of previour work, we observe that authors apply several different datasets to validate their proposed prediction models. We observe that several authors have used only a limited number of software systems to investigate the relationships between fault proneness and object-oriented metrics. Hence we beleive that due to experiments being conducted on limited dataset, it may not be clear whether a particular conclusion could be generalized to other software systems. However, in our study presented this paper, we have considered 45 real-life datasets from the PROMISE
repository \footnote{\url{http://openscience.us/repo/defect/}\label{rf1}}. PROMIS repository is one if the largest repositories for software engineering research data. The dataset on PROMISE repository is publicly available and it is very easy to find research dataset on the repository. By conducting experiments on dataset form PPROMISE repository, we make our experiments easily replicable for other researchers. Table \ref{Faulty} shows the percentage of faulty classes in each project in the PROMISE repository used in our experiments. 

The experimental dataset consists of several popular and widely used applications. Apache Ant\footnote{\url{https://ant.apache.org/}} is a Java library and command-line tool which is used for building Java applications and provides a number of built-in tasks allowing developers to compile, assemble, test and run Java applications. Log4j\footnote{\url{https://logging.apache.org/}} is a popular logging package for Java and is distributed under the Apache Software License. Apache Lucene\footnote{\url{https://lucene.apache.org/}} is a high-performance, full-featured text search engine library written entirely in Java. jEdit\footnote{\url{http://www.jedit.org/}} is a text editor written in Java for which hundreds of macros and plugins available. 
\begin{table}[h]
	\centering
	\caption{List of projects used in our experiments, number of classes and percentage of faulty classes}
	\label{Faulty}
	\renewcommand{\arraystretch}{1.1}
	\resizebox{8.5cm}{!}{
		\begin{tabular}{|l|*{4}{c|}r}
			
			\hline
			
			\textbf{Id} & 	\textbf{Project}  &  \textbf{No. of class} & \textbf{No. of Faulty class} &  \textbf{Faulty (\%)} \\
			\hline
			
			D1 & ant-1.3 & 125 & 20 & 16  \\ \hline
			D2 & ant-1.4 & 178 & 40 & 22.47  \\ \hline
			D3 & ant-1.5 & 293 & 32 & 10.92  \\ \hline
			D4 & ant-1.6 & 351 & 92 & 26.21  \\ \hline
			D5 & ant-1.7 & 745 & 166 & 22.28  \\ \hline
			D6 & arc & 234 & 27 & 11.54  \\ \hline
			D7 & berek & 43 & 16 & 37.21  \\ \hline
			D8 & camel-1.0 & 339 & 13 & 3.83  \\ \hline
			D9 & camel-1.2 & 608 & 216 & 35.53  \\ \hline
			D10 & camel-1.4 & 872 & 145 & 16.63  \\ \hline
			D11 & camel-1.6 & 965 & 188 & 19.48  \\ \hline
			D12 & e-learning & 64 & 5 & 7.81  \\ \hline
			
			D13 & ivy-1.1 & 111 & 63 & 56.76  \\ \hline
			D14 & ivy-1.4 & 241 & 16 & 6.64  \\ \hline
			D15 & ivy-2.0 & 352 & 40 & 11.36  \\ \hline
			D16 & jedit-3.2 & 272 & 90 & 33.09  \\ \hline
			D17 & jedit-4.0 & 306 & 75 & 24.51  \\ \hline
			D18 & jedit-4.1 & 312 & 79 & 25.32  \\ \hline
			D19 & jedit-4.2 & 367 & 48 & 13.08  \\ \hline
			
			D20 & kalkulator & 27 & 6 & 22.22  \\ \hline
			D21 & log4j-1.0 & 135 & 34 & 25.19  \\ \hline
			D22 & log4j-1.1 & 109 & 37 & 33.94  \\ \hline
			D23 & log4j-1.2 & 205 & 189 & 92.2  \\ \hline
			D24 & lucene-2.0 & 195 & 91 & 46.67  \\ \hline
			D25 & lucene-2.2 & 247 & 144 & 58.3  \\ \hline
			D26 & lucene-2.4 & 340 & 203 & 59.71  \\ \hline
			
			D27 & pdftranslator & 33 & 15 & 45.45  \\ \hline
			D28 & prop-1 & 18471 & 2738 & 14.82  \\ \hline
			D29 & prop-2 & 23014 & 2431 & 10.56  \\ \hline
			D30 & prop-3 & 10274 & 1180 & 11.49  \\ \hline
			D31 & prop-4 & 8718 & 840 & 9.64  \\ \hline
			D32 & prop-5 & 8516 & 1299 & 15.25  \\ \hline
			D33 & prop-6 & 660 & 66 & 10  \\ \hline
			D34 & redaktor & 176 & 27 & 15.34  \\ \hline
			D35 & serapion & 45 & 9 & 20  \\ \hline
			
			D36 & synapse-1.0 & 157 & 16 & 10.19  \\ \hline
			D37 & synapse-1.1 & 222 & 60 & 27.03  \\ \hline
			D38 & synapse-1.2 & 256 & 86 & 33.59  \\ \hline

			D39 & termoproject & 42 & 13 & 30.95  \\ \hline
			
			D40 & velocity-1.5 & 214 & 142 & 66.36  \\ \hline
			D41 & velocity-1.6 & 229 & 78 & 34.06  \\ \hline
			
			D42 & xerces-1.2 & 440 & 71 & 16.14  \\ \hline
			D43 & xerces-1.3 & 453 & 69 & 15.23  \\ \hline
			D44 & xerces-1.4 & 588 & 437 & 74.32  \\ \hline
			D45 & xerces-init & 162 & 77 & 47.53  \\ \hline

		\end{tabular}}
	\end{table}

\subsection{Research Questions}
Based on our analysis of several techniques on fault prediction as well as literature review studies, we frame the following research questions as gaps and contribution to the body of knowledge on software fault prediction: \\
\begin{itemize}
\item [RQ1] \textit{Do source code  metrics predict faulty or non faulty classes?} \\

The objective of this question is to test the relationship between each source code metric and fault proneness. In this research question, t-test analysis is used to test the statistical significance between faulty and non-faulty group metrics. T-test statistical significance testing is done by analyzing the means and distribution of faulty and non-faulty group metrics.\\ 

\item [RQ2] \textit{Can the selected set of source code metrics better predict whether a class is faulty or not?} \\
	
In this research question, our aim to evaluate the performance of the selected sets of metrics. In this study, two steps (i.e. t-test  and forward stepwise selection procedure) are considered to identify subsets of object-oriented software metrics (from ) that are more capable of predicting whether class is faulty or not.\\

\item [RQ3] \textit{Which fault prediction technique is a most suitable one for this purpose among all?} \\

This question helps to investigate the performance of different types of fault-prediction techniques. To address this question, three ensemble methods are considered for developing a fault prediction models in order to achieve the best performance.\\

\item [RQ4] \textit{For any given software product, is performing fault prediction analyses economically effective?} \\

This question investigates the effectiveness of different fault prediction techniques. To address this issue, a cost evaluation framework is proposed which performs cost based analysis for mis-classification of faults. \\
\end{itemize}

\section{Source Code Metrics Validation Framework}
\label{evm}
Researchers' uses different set of source code metrics as input to develop a model for predicting whether class is faulty or not \cite{chen2009empirical} \cite{henry}\cite{Basili}.  
This shows that the performance of fault prediction model depends on the software metrics which have been considered as inputs to develop a model. Selection of a suitable set of features is an important data pre-processing task in different applications of data mining and machine learning \cite{forman2003extensive}\cite{furlanello2003entropy}\cite{doraisamy2008study}. In this paper, a source code metrics validation framework has been proposed for validating the metrics and identifying suitable set of source code metrics with an aim to reduce irrelevant metrics and improve the performance of the fault prediction model. Irrelevant source code metrics are those features with very low predictive or discriminatory power with respect to the target class. The proposed framework  is applied to 45 real-life datasets taken from the PROMISE repository \footnote{http://openscience.us/repo/}. Finally, we have validated the framework by comparing the performance of the models developed using a selected set of source code metrics with the performance of those developed using original dataset.

\begin{figure}[h]

	\centering
	\begin{tikzpicture}[scale=0.8]

	\node[text width=2cm, text badly centered,draw,rounded corners=0.2cm, very thick, font=\scriptsize] at (5,26.25) (b) {Normalization of Data };
	
	\node[text width=1.5cm, text badly centered,draw,rounded corners=0.2cm, very thick, font=\scriptsize] at (8.5,26.25) (c) {t-test Analysis};

	\node[text width=2.5cm, text badly centered,draw,rounded corners=0.2cm, very thick, font=\scriptsize] at (8.5,24.25) (d) {Multivariate Linear Regression Stepwise Forward Selection};

	\node[text  width=3.5cm, text badly centered,draw=none,fill=blue!20,rounded corners=0.2cm,fill=none,below,font=\scriptsize] at (1,29)  {Data set containing software metrics and fault in software modules};

	\node[text  width=2.50cm, text badly centered,draw=none,fill=blue!20,rounded corners=0.2cm,fill=none,below,font=\scriptsize] at (5,28.5)  {Metrics are normalized over the range between $0$ to $1$ i.e., [0, 1]};
	
	\node[text  width=2.50cm, text badly centered,draw=none,fill=blue!20,rounded corners=0.2cm,fill=none,below,font=\scriptsize] at (8.5,28.5)  {pre-processing step: selection of metrics without involving learning algorithm};

	\node[text  width=3.0cm, text badly centered,draw=none,fill=blue!20,rounded corners=0.2cm,fill=none,below,font=\scriptsize] at (4.7,24.9)  {Feature selection step: This analysis search right set of metrics for fault prediction. };

	\draw[->, very thick] (2,26.25) to (b);
	\draw[->, very thick] (b) to (c);
	\draw[->, very thick] (c) to (d);

	\node[cylinder,draw=black,thick,aspect=0.7,
	minimum height=1.4cm,minimum width=2.5cm,
	shape border rotate=90,
	cylinder uses custom fill]  at (1,26)
	(a) {Data Set};];
	
	\end{tikzpicture}
	\caption{Proposed Framework of Software Metrics Validation}
	\label{rs}	
	
\end{figure}
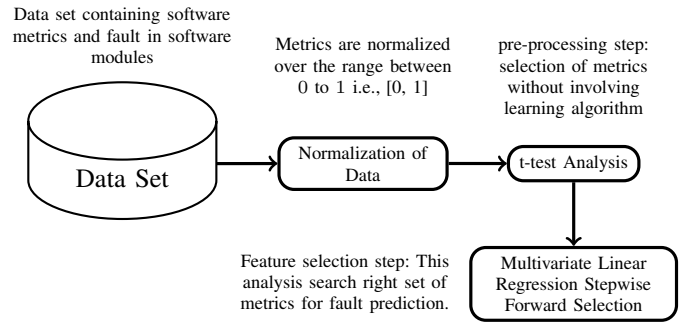

Figure \ref{rs} shows the detail steps of the proposed software metrics validation framework. Our proposed approach is a multi-step process. The objective of this framework is to first investigate whether these source code metrics are significant predictors of fault proneness without involving any learning algorithm. After identifying all significant source code metrics, the wrapper approach is employed for identifying the right set of source code metrics. It uses the performance of the chosen learning algorithm  to evaluate each candidate feature subset. In this experiment, linear discriminant analysis is considered as a classification algorithm.

\begin{itemize}
	\item [i.] \textit{Data Set:} The requisite fault data of 45 projects listed in Table \ref{Faulty} are taken from tera-PROMISE Repository. The data set containing fault information and twenty software source code metrics are also included.
	
	\item [ii.] \textit{Normalization of Data:} All source code metrics are normalized over the range between $0$ to $1$ using Min-Max normalization technique. The normalization of source code metrics is required to adjust the defined range of metrics. Hence, before applying the machine learning algorithm and subsequent t-test analysis, the input metrics are normalized or standardized using min-max scaling.
	\item [iii.] \textit{Filter Approach:}  Filter approach is usually used as a pre-processing step to remove insignificant features. We have considered the characteristics of features to select significant sets of features without involving a learning algorithm. T-test  has been employed  to remove insignificant features. The objective of this step is to test the relationship between each source code metric and fault proneness. In this study, t-test analysis is used to test the statistical significance between faulty and non-faulty group metrics. In 2-class problems (faulty class and non faulty classes), test of null hypothesis ($H0$) means that the two populations are not equal; on other words, there is a significant difference between their mean value and both features are different. It further implies that the metrics affect the fault prediction result. Hence, these metrics have been considered and those having no significant difference between their mean value are rejected. Therefore, it is necessary to accept the null hypothesis ($H0$) and reject the alternate hypothesis. Here, a $t-test$ on each metric is applied and compared with their corresponding $P-value$ for each metric as a measure of how effectively it separates the groups. The metrics having $P-value$ smaller than 0.05 have strong discrimination powers. 
\item [iv.] \textit{Wrapper Approach:} The above filter approach does not consider the interaction between source code metrics; hence, it may be possible that the selected metrics using filter approach also contain redundant information. In this work, multivariate linear regression stepwise forward selection is considered in a wrapper fashion to compute an optimal set of source code metrics. 
\end{itemize}
\section{Cost Analysis Model}
\label{coana}
In this section, we describe the construction of our proposed cost evaluation model. The cost evaluation model accounts for the realistic costs incurred to remove a fault or defect in the system and computes the estimated fault removal cost for a specific fault prediction technique based on the ideas proposed by Wagner et al. (\cite{a35}). We make certain assumptions and define the constraints in designing this cost evaluation model. The assumptions and constraints are listed as follows:

\begin{itemize}
	\item[i.] Different testing phases account for varying fault removal cost.
	\item[ii.] It is not practically possible to completely detect all faults within one testing phase.
	\item[iii.] It is not practically possible to perform unit testing on all classes in a system.
\end{itemize}

The normalized fault removal cost approach suggested by Wagner \textit{et al.} (\cite{a35}) is applied in our work to formulate the proposed cost evaluation model. The fault removal cost varies because various projects are developed on different platforms following organization standards and practises which varies across projects. The normalized fault removal costs are summarized in Table \ref{NormCost}. The fault identification efficiencies for different testing phases used in our work are inspired from the study by Jones et al. (\cite{a20}). The efficiencies of testing phases are summarized in Table \ref{FaultIde_DiffTestPhases}. Wilde \textit{et al.} (\cite{a36}) stated that more than $50\%$ of the classes are usually very small in size, and hence performing unit testing on these classes may not be very much helpful.

\begin{table}[h!]
	\caption{Removal costs of test techniques (in staff hour per defects)}
	\centering
	\begin{tabular}{|l|*{5}{c|}r}
		
		\hline
		\textbf{Type} & \textbf{Min} & \textbf{Max} & \textbf{Mean} & \textbf{Median} \\ \hline
		Unit ($C_{u}$)& 1.5 & 6 & 3.46 & 2.5 \\ \hline
		Integration ($C_{i}$) & 3.06 & 9.5 & 5.42 & 4.55\\ \hline
		System ($C_{s}$) & 2.82 & 20 & 8.37 & 6.2 \\ \hline
		Field ($C_{f}$) & 3.9 & 66.6 & 27.24 & 27 \\
		\hline
	\end{tabular}
	
	\label{NormCost}

	\caption{Fault identification efficiencies - test phases}
	\centering
	\begin{tabular}{|l|*{4}{c|}r}
		\hline
		\textbf{Type} & \textbf{Min} & \textbf{Max } & \textbf{Median} \\ \hline
		Unit ($\delta_{u}$) & 0.1 & 0.5 & 0.25  \\ \hline
		Integration ($\delta_{i}$) & 0.25 & 0.60 & 0.45 \\ \hline
		System ($\delta_{s}$) & 0.25 & 0.65 & 0.5 \\ \hline
	\end{tabular}
	\label{FaultIde_DiffTestPhases}
	
\end{table}

The formulations of $E_{cost}$, $T_{cost}$ and the $NE_{cost}$ of the proposed cost based evaluation framework are presented int he below section. The mathematical notations used in our framework are described below:

\begin{itemize}
	\item [i.] $C_{i}$: Initial setup cost of used fault-prediction technique, $C_{u}$, $C_{i}$, $C_{s}$, and $C_{f}$ are the normalized fault removal cost in unit, integration, system, and field testing respectively. $M_{p}$: percentage of classes unit tested.
	\item [ii.] $\delta_{u}$, $\delta_{i}$ and $\delta_{s}$ are the fault identification efficiency of unit, integration, and system testing respectively.
	\item [iii.] $FC$ and $TC$ are the number of faulty modules and total number of modules in software projects respectively. $TN$, $FN$, $FP$, and $TP$ are the value of true negative, false negative, false positive, and true positive respectively.
\end{itemize}
\textbf{Estimated fault removal cost ($E_{cost}$)}: The series of steps involved in computing the estimated fault removal cost of the software system when fault prediction is performed ($E_{cost}$) are defined as follows:

\begin{itemize}
	\item [i.] Total number of faulty classes identified by the predictor are equal to the summation of true positive ($ TP $) and false positive ($ FP $) values. Hence, it is necessary to compute testing and verification cost at class level of granularity which indicates that the value of cost is equal to the cost of unit testing ($C_{u}$). The total cost on unit testing of software system is defined as:
	\begin{equation}
	\label{eust}
	Cost_{unit}=(TP+FP)*C_{u}
	\end{equation}
	\item [ii.] Since it impractical to detect all fault within a specific testing phase, there is a possibility that some of the correctly predicted faulty classes remain undetected in unit testing. Furthermore, there is a possibility that these faulty classes which were predicted as non-faulty classes (number of false negative ($ FN $)), are identified by the predictor in the later phases of testing, such as integration($C_{i}$), system, and field testing. The fault removal cost in integration, system, and field testing is computed as follows:
	\begin{multline}
	\label{init}
	Cost_{Integration}= C_{i}*\delta_{i}*(FN+TP*\\(1-\delta_{u}))
	\end{multline}
	\begin{multline}
	\label{sust}
	Cost_{system}= \delta_{s}*C_{s}*((1-\delta_{i})*(TP*\\(1-\delta_{u})+FN))
	\end{multline}
	\begin{multline}
	\label{fust}
	Cost_{field}= (1-\delta_{s})*C_{f}*((1-\delta_{i})*(TP*\\
	(1-\delta_{u})+FN))
	\end{multline}
	
	\item [iii.] The estimated overall fault removal cost can be determined as:
	\begin{multline}
	\label{EqnEcost}
	Ecost=C_{i}+C_{u}*(FP+TP) + \delta_{i}*C_{i}*(TP*\\(1- \delta_{u})+FN) +
	\delta_{s}*C_{s}*((1-\delta_{i})*(TP*(1-\delta_{u})+ \\FN))+ 
	(1-\delta_{s})*C_{f}*((1-\delta_{i})*(TP*(1-\delta_{u})+FN))
	\end{multline}

\end{itemize}
\textbf{Estimated testing cost ($T_{cost}$):} The list of steps to compute the estimated fault removal cost of the software system without using fault prediction approach ($T_{cost}$) is defined as follows:
\begin{itemize}
	\item [i.] In testing phase, if fault prediction analysis is not conducted, then the testing team often performs unit testing on all the classes. Therefore, total unit testing may be computed as:
	\begin{equation}
	\label{eust1}
	Cost_{unit}= M_{p}*C_{u}*TC
	\end{equation}
	\item [ii.] Furthermore, there is a likelihood that some of the faulty classes that remain undetected in unit testing may later be identified in integration, system, and field testing phases.  The total integration, system, and field testing cost is calculated as:
	\begin{equation}
	\label{sust11}
	Cost_{integration}= \delta_{i}*C_{i}*(1-\delta_{u})*FC
	\end{equation}
	\begin{equation}
	\label{sust12}
	Cost_{system}= \delta_{s}*C_{s}*((1-\delta_{i})*(1-\delta_{u})*FC)
	\end{equation}
	
	\begin{equation}
	\label{fust1}
	Cost_{field}= (1-\delta_{s})*C_{f}*((1-\delta_{i})*(1-\delta_{u})*FC)
	\end{equation}
	
	\item [iii.] The estimated overall fault removal cost  without the use of fault prediction can be determined by using following equation as: 
	\begin{multline}
	\label{EqnTcost}
	Tcost=M_{p}*C_{u}*TC+ \delta_{i}*C_{i}*(1-\delta_{u})*FC  \\ +\delta_{s}*C_{s}*((1-\delta_{i})*(1-\delta_{u})*FC+ (1-\delta_{s})\\*C_{f}*((1-\delta_{i})*(1-\delta_{u})*FC)
	\end{multline}

\end{itemize}
	
	\textbf{Normalized fault removal cost ($NE_{cost}$)} : Normalized fault removal cost ($\frac{E_{cost}}{T_{cost}}$) and its interpretation can be modeled as:	
	\begin{equation}
	\label{EqnNEcost}
	\text{If the value of }	NEcost = \left\{
	\begin{array}{ll}
	< 1, \text{then application of}\\\text{fault prediction is useful}
	
	\\
	
	\\
	=> 1, \text{then application of }\\\text{testing methodlogies may} \\\text{be helpful}
	\end{array}
	\right.
	\end{equation}

\section{Experimental setup}
\label{es}
\begin{figure*}[t]

	\centering
	\begin{tikzpicture}[scale=0.8]
	
	\draw[-, very thick] (3.5,23.90) to (6.5,23.90);
	\draw[-, very thick] (3.5,23.90) to (3.5,31);
	\draw[-, very thick] (3.5,31) to (6.5,31);
	\draw[-, very thick] (6.5,31) to (6.5,23.90);

	\node[text width=1.85cm, text badly centered,draw,fill=none, very thick, font=\scriptsize] at (5,30.0) (j1) {All Metrics (AM)};	
	\node[text width=1.85cm, text badly centered,draw,fill=none, very thick, font=\scriptsize] at (5,25.7500) (j1) {Selected Metrics using proposed framework};
	
	\node[text  width=3.0cm, text badly centered,draw=none,rounded corners=0.2cm,fill=none,below,font=\scriptsize] at (5.0,23.7)  {Set of source code metrics};
	
	\node[text width=1.85cm, text badly centered,draw,fill=none, very thick, font=\scriptsize] at (9.5,30.5) (j1) {LOGR};	
	\node[text width=1.85cm, text badly centered,draw,fill=none, very thick, font=\scriptsize] at (9.5,29.7) (j1) {ANN};	
	\node[text width=1.85cm, text badly centered,draw,fill=none, very thick, font=\scriptsize] at (9.5,28.9) (j1) {RBFN-RAN};	
	\node[text width=1.85cm, text badly centered,draw,fill=none, very thick, font=\scriptsize] at (9.5,28.100) (j1) {RBFN-FCM};	
	\node[text width=1.85cm, text badly centered,draw,fill=none, very thick, font=\scriptsize] at (9.5,27.300) (j1) {RBFN-KCM};	
		\node[text width=1.85cm, text badly centered,draw,fill=none, very thick, font=\scriptsize] at (9.5,26.500) (j1) {RBFN-KCM};	
			\node[text width=1.85cm, text badly centered,draw,fill=none, very thick, font=\scriptsize] at (9.5,25.700) (j1) {RBFN-KCM};	
				\node[text width=1.85cm, text badly centered,draw,fill=none, very thick, font=\scriptsize] at (9.5,24.900) (j1) {RBFN-KCM};

	\draw[-, very thick] (8,23.90) to (11,23.90);
	\draw[-, very thick] (8,23.90) to (8,31);
	\draw[-, very thick] (8,31) to (11,31);
	\draw[-, very thick] (11,31) to (11,23.90);
	
	\draw[-, very thick] (13,23.90) to (16,23.90);
	\draw[-, very thick] (13,23.90) to (13,31);
	\draw[-, very thick] (13,31) to (16,31);
	\draw[-, very thick] (16,31) to (16,23.90);
	
	\node[text width=1.85cm, text badly centered,draw,fill=none, very thick, font=\scriptsize] at (14.5,30.2) (j1) {Performance Evaluation};	
	
	\node[text width=1.85cm, text badly centered,draw,fill=none, very thick, font=\scriptsize] at (14.5,27.500) (j1) {Statistical Test};
	
	\node[text width=1.85cm, text badly centered,draw,fill=none, very thick, font=\scriptsize] at (14.5,25.100) (j1) {Model comparsion};
	
	\node[text  width=3.5cm, text badly centered,draw=none,rounded corners=0.2cm,fill=none,below,font=\scriptsize] at (9.5,23.7)  {Classification Techniques with 10-fold cross validation};
	
	\node[text width=1.85cm, text badly centered,draw,fill=none, very thick, font=\scriptsize] at (19.5,30.2) (j1) {$E_{cost}$ Evaluation };	
	\node[text width=1.85cm, text badly centered,draw,fill=none, very thick, font=\scriptsize] at (19.5,27.7500) (j1) {$T_{cost}$ Evaluation};	
	\node[text width=1.85cm, text badly centered,draw,fill=none, very thick, font=\scriptsize] at (19.5,25.100) (j1) {$NE_{cost}$ Evaluation};
	
	\node[text  width=3.5cm, text badly centered,draw=none,rounded corners=0.2cm,fill=none,below,font=\scriptsize] at (14.5,23.7)  {Validation of developed models};
	
	\node[text  width=3.5cm, text badly centered,draw=none,rounded corners=0.2cm,fill=none,below,font=\scriptsize] at (19.5,23.7)  {Cost Analysis};
	
	\draw[-, very thick] (18,23.90) to (21,23.90);
	\draw[-, very thick] (18,23.90) to (18,31);
	\draw[-, very thick] (18,31) to (21,31);
	\draw[-, very thick] (21,31) to (21,23.90);

	\draw[-, very thick] (11,28) to (12.5,28);
	\draw[-, very thick] (11,27) to (12.5,27);
	\draw[-, very thick] (12.5,28.5) to (13,27.5);
	\draw[-, very thick] (12.5,26.5) to (13,27.5);
	\draw[-, very thick] (12.5,28.5) to (12.5,28);
	\draw[-, very thick] (12.5,26.5) to (12.5,27);
	
	\draw[-, very thick] (16,28) to (17.5,28);
	\draw[-, very thick] (16,27) to (17.5,27);
	\draw[-, very thick] (17.5,28.5) to (18,27.5);
	\draw[-, very thick] (17.5,26.5) to (18,27.5);
	\draw[-, very thick] (17.5,28.5) to (17.5,28);
	\draw[-, very thick] (17.5,26.5) to (17.5,27);
	
	
	\draw[-, very thick] (6.5,28) to (7.5,28);
	\draw[-, very thick] (6.5,27) to (7.5,27);
	\draw[-, very thick] (7.5,28.5) to (8.0,27.5);
	\draw[-, very thick] (7.5,26.5) to (8.0,27.5);
	\draw[-, very thick] (7.5,28.5) to (7.5,28);
	\draw[-, very thick] (7.5,26.5) to (7.5,27);
	
	\draw[-, very thick] (2,28) to (3,28);
	\draw[-, very thick] (2,27) to (3,27);
	\draw[-, very thick] (3,28.5) to (3.5,27.5);
	\draw[-, very thick] (3,26.5) to (3.5,27.5);
	\draw[-, very thick] (3,28.5) to (3,28);
	\draw[-, very thick] (3,26.5) to (3,27);

	\node[cylinder,draw=black,thick,aspect=0.7,
	minimum height=1.0cm,minimum width=2.0cm,
	shape border rotate=90,
	cylinder uses custom fill,
	]  at (1.0,27.1)
	(g) {Data Set};];	
	\end{tikzpicture}
	\caption{Framework of Proposed work}
	\label{fig1.2}
	
\end{figure*}
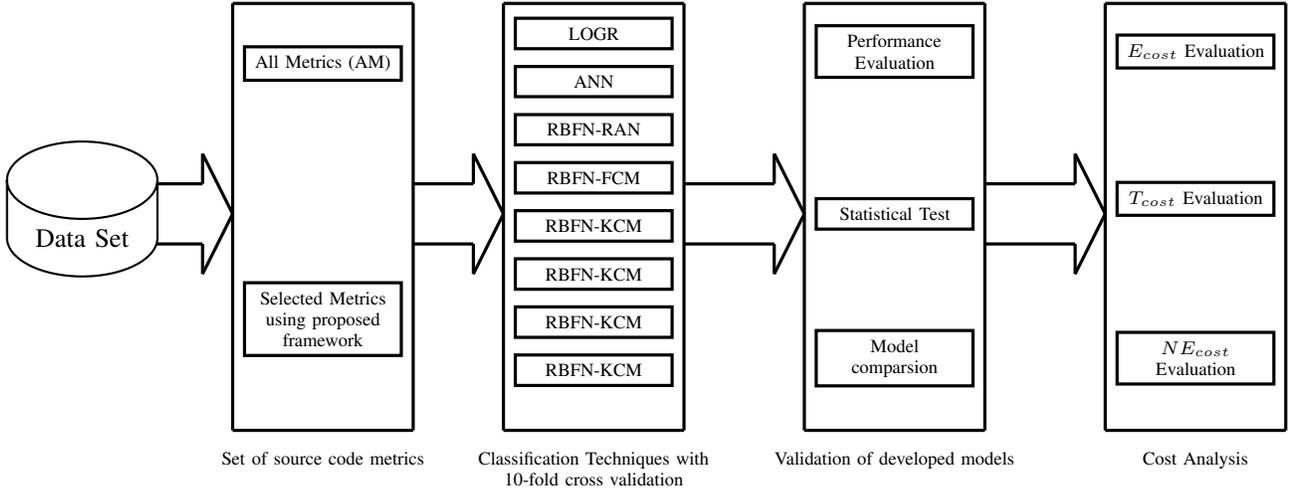

In this section, we present the experimental setup of our developed fault prediction models. Figure \ref{fig1.2} shows the work flow of the proposed work. In this study, we perform the following steps to develop fault prediction models:\\

\noindent\textbf{Metrics Selection:} Selection of suitable set of source code metrics using proposed source code metrics validation framework.\\

\noindent\textbf{Prediction Model:} Development of prediction model by considering source code metrics as input to predict fault proneness.\\

\noindent\textbf{Performance Measures:} Selection of performance measures that can used to evaluate the predictive capability of fault prediction models. \\

\noindent\textbf{Validation Methods:} Use of efficient validation methods to determine the true predictive applicability of the developed models.\\

\noindent\textbf{Statistical Tests:} Selection of appropriate statistical tests to determine the superiority of one prediction technique over the other prediction techniques and also determine the superiority of one set of source code metrics over the other sets.\\

\noindent\textbf{Experimental Validation:} Validation of developed fault prediction models using proposed cost analysis framework.

\subsection{Selection of source code metrics}
In this work, 20 source code metrics have been used for fault prediction. It is very essential to remove irrelevant and unimportant source code metrics out of these source code metrics so that only  uncorrelated and relevant source code metrics are included in the construction of fault prediction models. In order to achieve this objective, we have proposed source code metrics validation framework as described in section \ref{evm}. This proposed framework is used for removing irrelevant source code metrics and select right set of metrics for fault prediction. 

\subsection{Classification Techniques}
In this work, we have carefully selected three  different ensemble methods. In ensemble of classification models, we have considered the outputs of all its individual constituent classification models where base learners are assigned a certain priority level in the each classification model and the final output is computed with the help of some combination rules. There are two types of ensemble methods,
\begin{itemize}
	\item \textbf{Homogeneous Ensemble Method}: In this method, all considered base learners, i.e. classification models, are of the same types, but each one has a randomly generated training set \cite{bahler2000methods}\cite{bian2007diversity}\cite{xu2015efspredictor}.
	\item \textbf{Heterogeneous Ensemble Method}: In this method, all considered base learners, i.e. classification models, are of different types \cite{bahler2000methods}\cite{bian2007diversity}\cite{xu2015efspredictor}.
\end{itemize}
The ensemble methods can be further categorized into two different groups based on combination rules. These categories are:
\begin{itemize}
	\item \textbf{Linear Ensemble Method}: In this method, the arbitrator combines the outputs of the base learners, i.e. classification models, in a linear fashion such as averaging, best in training, weighted averaging, etc.
	\item \textbf{Nonlinear Ensemble Method}: In this method, the output of the considered base learners, i.e. classification models, are fed into an arbitrator, which is a nonlinear prediction model such as neural network, Decision tree forest (DTF) etc.
\end{itemize}
In the present work, we have considered a heterogeneous ensemble method with three different combination rules (2 Linear, and 1 nonlinear). A detailed description of the ensemble methods used in this work are tabulated in Table \ref{esa}.

\begin{table}[h!]
	\centering
	\caption{Ensembles of Classification Models}
	\label{esa}
	\renewcommand{\arraystretch}{1.1}
	\resizebox{8.50cm}{!}{
		\begin{tabular}{l*{2}{c}r}
			\hline
			 \textbf{Base Learners} & \textbf{Combination Rules} \\ \hline
		LOGR, ANN, RBFN-RAN, RBFN-FCM, RBFM-KCM & Linear (best in training) \\ 
		 LOGR, ANN, RBFN-RAN, RBFN-FCM, RBFM-KCM & Linear (majority voting) \\
		LOGR, ANN, RBFN-RAN, RBFN-FCM, RBFM-KCM & Non-linear (DTF) \\ \hline
			
		\end{tabular}}
	\end{table}

\subsection{Base learners}
In this section, we briefly describe each base learners that were used in ensemble methods.

\subsubsection{Logistic regression analysis (LOGR)} 
Logistic regression is a statistical method for analyzing a dataset in which there are one or more independent variables that predicting the outcomes of dependent variable \cite{chidamber1991towards}. The logistic regression model is based on the following equation:

\begin{equation}\ \pi(x) = \frac{{e}^{{\alpha}_{0}+ \sum\limits_{i=1}^{p} \alpha_{i}*X_{i}}}{1+{e}^{{\alpha}_{0}+\sum\limits_{i=1}^{p} \alpha_{i}*X_{i}}} \end{equation}

$‘P’$ represents the number of independent variables.  $\pi$  represents the probability of fault  in the class during validation.
\subsubsection{Artificial Neural Network (ANN) model}
 In the present work, ANN is considered for developing fault prediction models. 
 The neural network can be represented as:
 \begin{equation}\ O' = f(W,I)\end{equation}
 
 where $I$ and $O^{'}$  are the input and desired output vectors. $W$ is the weight vector, whose $W$ value is updated in every iteration with an aim to reduce the value of mean square error (MSE). MSE is computed using the following equation:
 
 \begin{equation}
 \label{msef}
 MSE = \dfrac{1}{n} \sum_{i=1}^{n} (O'_{i}-O_{i})^{2}\end{equation}
where $O$, and $O^{'}$ are the actual and desired output values. In the present work, Gradient Descent method is considered for training the ANN model. The Gradient Descent (GD) method is used for updating the weights to minimize the output error \cite{Batti}. GD method uses the 1st order derivative of the total error function to find the minima in error space. It is represented using the following equation:

\begin{equation}
\label{grd}
G=\frac{\partial }{\partial  W} \big( E_{k} \big)=\frac{\partial}{\partial W} \big( \dfrac{1}{2} (O'_{k}-O_{k})^{2} \big)
\end{equation}
In each iteration, weight vector $W$ is updated using gradient vector $G$ \footnote{http://in.mathworks.com/help/nnet/ref/traingd.html}. Weighted vector $W$ is updated as:
\begin{equation}
\label{we_grad}W_{k+1}=-\alpha G_{k }=-\alpha\frac{\partial }{\partial  W} \big( E_{k} \big) \end{equation}

where  $W_{k+1}$ is the updated weight vector, $G_{k}$ is the gradient vector and $\alpha$ is the learning constant. We apply the cross-validation technique to compute the optimal value of $\alpha$.
\subsubsection{Radial Basis Function Neural Network (RBFN)} 
RBFN network \cite{zhang2016direct} is a popular alternative to the feed forward neural network, since it has simple network structure and fast training process \cite{zhang2016direct}.  
 In this analysis Hybrid RBFN model has been used for predicting software fault proneness. Basically RBFN model involves updating the centers and the weights of the model. In this paper, different approaches have been followed for updating centers and weight, details of which are mentioned below:
 \begin{itemize}
 	\item[a.] In the first approach, centers are generated randomly and weights are updated using Gradient Descent, is referred to as RBFN-RAN.
 	
 	\item[b.] In the second instance, centers are identified using K-means clustering and weights are updated using Gradient Descent, and this approach is termed as RBFN-KMC.

 	\item[c.] In the third approach, centers are identified using Fuzzy C-mean clustering technique and weights are updated using Gradient Descent, and is referred to as RBFN-FCM.	
 \end{itemize}

\subsection{Best Training Ensemble (BTE)}
Best Training Ensemble (BTE) method takes the advantage of the fact that each classifier has a different performance across the used dataset partitions. Amongst these, we select the best model in the training dataset based on certain performance parameters. 
\subsection{Majority Voting Ensemble (MVE) Method}
In Majority Voting Ensemble (MVE) method, we have considered the output of each classifier on the test data, and the ensemble output ($E_{out}$) is the majority category classified by the base classifier.
\subsection{Nonlinear Ensemble Decision Tree Forest (NDTF)}
In nonlinear ensemble, we have considered the output corresponding to the training data of the base learner as the input to train the non-linear ensemble model.  The trained non-linear ensemble model uses the output corresponding to the testing data of the base learner to make a final prediction on the test set. In this study, we have considered Decision tree forest (DTF) as a classifier for non-linear ensemble. The concept of DTF was proposed by Breiman in 2001. It is a collection of different decision trees where the result of each tree is combined to make a final decision. 

\subsection{Performance Parameter}
In order to evaluate the fault prediction model, various performance parameters need to be analyzed which indicate the effectiveness of the developed fault prediction models. In this work, we have considered  two different performance parameters: accuracy and F-Measure. These parameters are computed using values of various elements in a confusion matrix as shown in Table \ref{conf}. 
\begin{table}[h!]
	\centering
	\caption{Confusion matrix to classify a class as faulty and not-faulty}
	\label{conf}
	\begin{tabular}{|l|*{3}{c|}r}
		\hline
		& \textbf{Non Faulty }  & \textbf{Faulty} \\
		\hline
		
		Non Faulty	& $N_{NF->NF}$ & $N_{NF->F}$\\
		\hline
		Faulty	& $N_{F->NF}$ & $N_{F->F}$ \\ 
		\hline
	\end{tabular}
\end{table}

\begin{equation}
	Accuracy=\frac{N_{NF->NF} + N_{F->F}}{N_{classes}}
	\end{equation}

	\begin{multline}
	F-Measure=\dfrac{2*Precision*Recall}{Precision+Recall}=\\
	\frac{2*N_{F->F}}{2*N_{F->F}+N_{NF->F} + N_{F->NF}}
	\end{multline}

\subsection{Validation method}
The objective of this work is to apply our developed model to predict faulty classes for  future releases and unseen similar natured projects. Hence, it is necessary to validate the developed fault prediction model on different data set from which they are trained. In this experiment, we have considered 10-fold cross validation to validate the proposed fault prediction model. Cross-validation is a statistical learning method, being used to evaluate and compare the models  by partitioning the data into two portions. One portion of the divided set is used to train or learn the model and the rest of the data is used to validate the model, based on training.  
\subsection{Statistical tests} 
In order to statistically analyze the results, we have considered pairwise Wilcoxon signed rank test. We conducted this test to determine which of prediction methods and feature selection techniques work better or all have performed equally well. We have analyzed all results at 0.05 significance level.
\subsection{Validation of Developed Fault Prediction Model}  
Finally these developed models are validated using proposed cost analysis framework.

\section{Experimental results}
\label{er}
\subsection{Source Code Metrics Validation}
\begin{figure*}[h!]
	\centering
	\includegraphics[width=17cm, height=6.5cm]{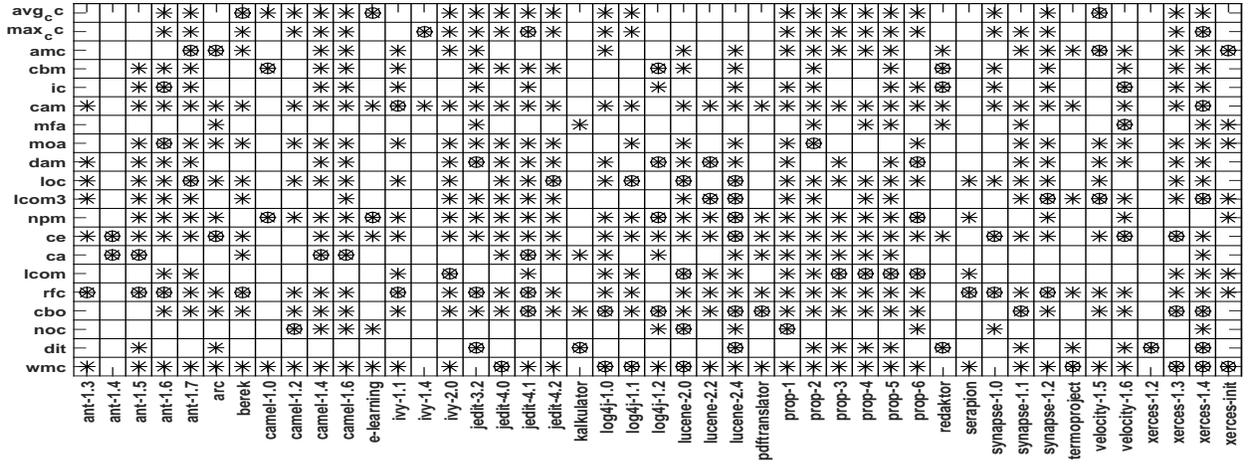}
	\caption{Selected Set of Metrics}
	\label{sm}
\end{figure*}

This section presents a detailed description of selection the right set of metrics for fault prediction. We started the statistical analysis with 20 source code object-oriented metrics. Figure \ref{sm} shows the selected set of metrics in each step for all 45 projects. For the purpose of simplicity, the graphs are represented using four different symbols as described below:
\begin{itemize}
	\item Empty circle ($\mathlarger{\circ}$): source code metrics selected after t-test analysis; and
 \item Circle with star ($\mathlarger{\mathlarger{\mathlarger{\mathlarger{\circ}}}}$\hspace{-1.250em} $\mathlarger{\mathlarger{\mathlarger{\ast}}}$): source code metrics selected after t-test and MLR stepwise forward selection method.
\end{itemize} 
From Figure \ref{sm}, it is observed that wmc, cbo	rfc, lcom, ca are commonly referred as relevant metrics to the fault of classes in most of the projects. 
\begin{figure}[h]
	\centering
	\subfloat[Accuracy \label{accuracy}]{
		
		\includegraphics[width=8.5cm, height=4.5cm]{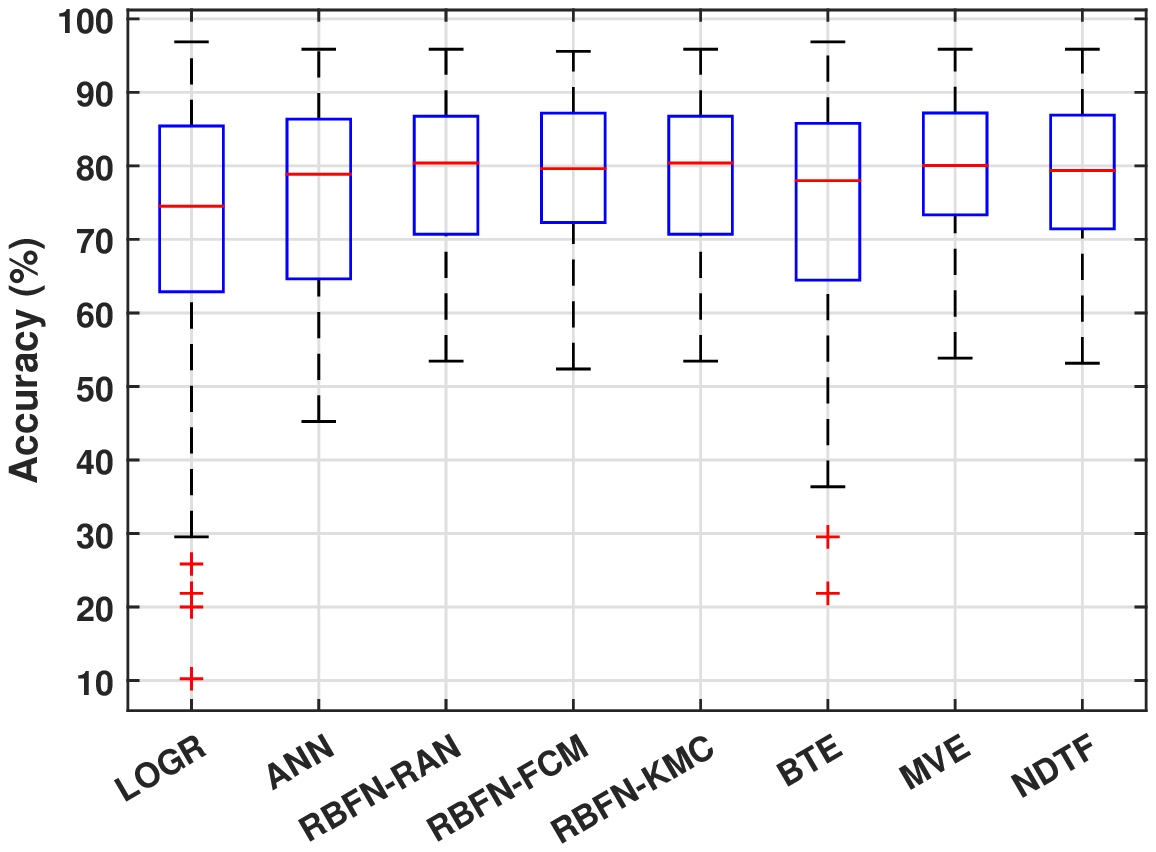} }
	
	\subfloat[F-Measure\label{fmeasure}]{
		
		\includegraphics[width=8.5cm, height=4.5cm]{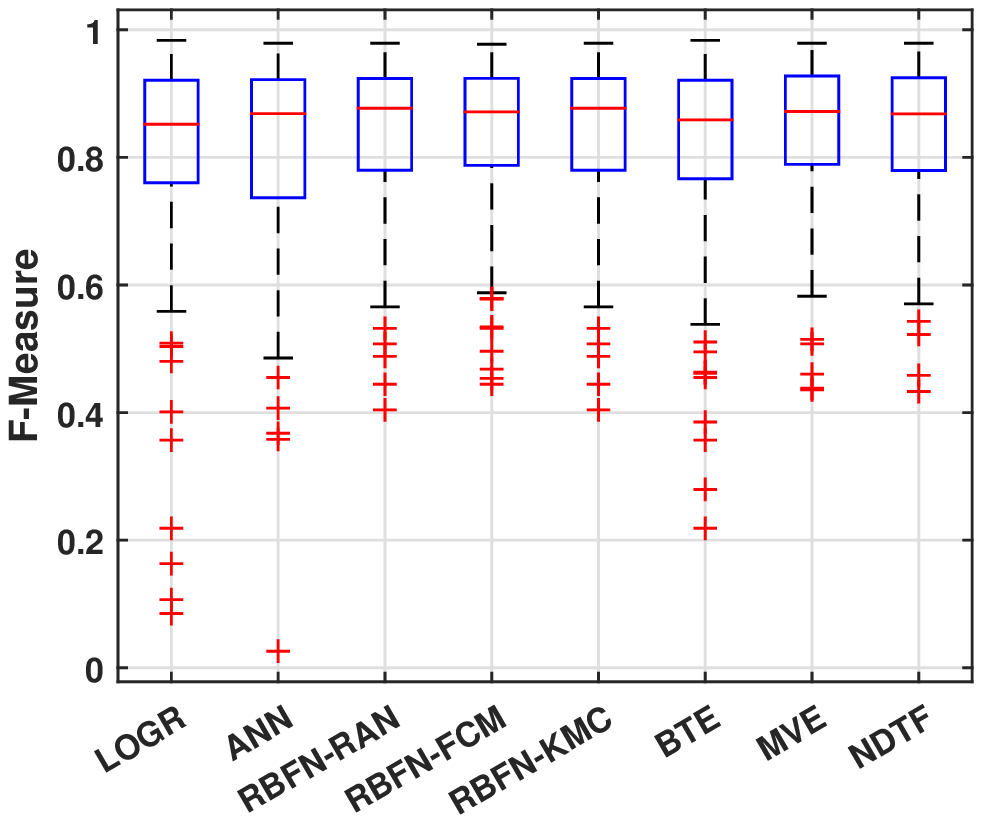} } 
	\caption{Classification Techniques}
	\label{perclass}
\end{figure}

\begin{figure}[h]
	\centering
	\subfloat[Accuracy \label{accuracysm}]{
		
		\includegraphics[width=3.5cm, height=4.5cm]{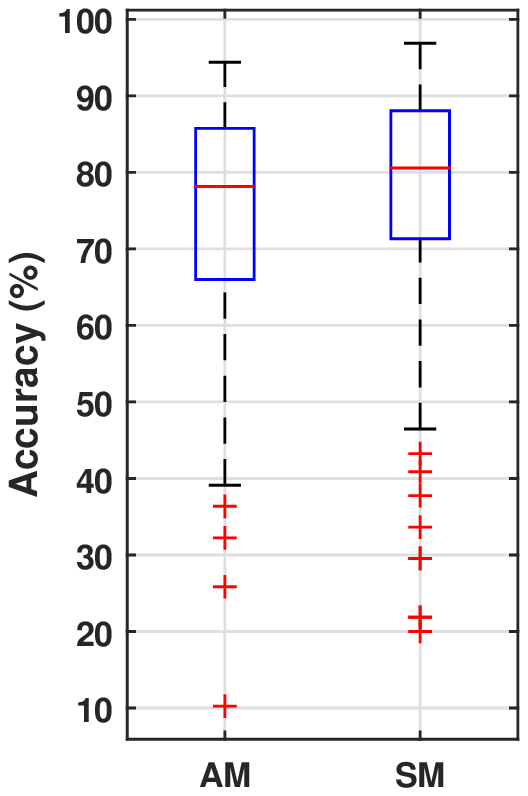} }
	\subfloat[F-Measure\label{fmeasuresm}]{
		
		\includegraphics[width=3.5cm, height=4.5cm]{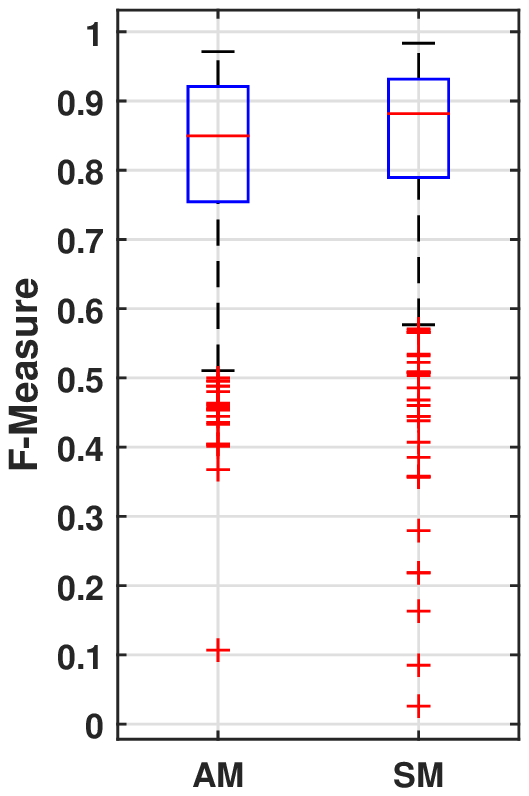} } 
	\caption{Selected set of source code metrics}
	\label{perclasssm}
\end{figure}
\begin{table}[h!]
	\caption{t-test}
	\label{tst}
	\centering
	\subfloat[Training Methods \label{testcm}]{
		\centering
		\renewcommand{\arraystretch}{1.1}
		\resizebox{8.5cm}{!}{
			\begin{tabular}{|l|*{9}{c|}r}
				\hline
					\multicolumn{9}{|c|}{\textbf{Accuracy}} \\ \hline
				  \multicolumn{9}{|c|}{\textbf{Mean}}	\\ \hline		
& \textbf{LOGR} &\textbf{ ANN} & \textbf{RBFN-RAN} & \textbf{RBFN-FCM} & \textbf{RBFN-KCM }& \textbf{BTE} & \textbf{MVE} & \textbf{NDTF } \\ \hline
LOGR & 0.00 & -6.46 & -8.78 & -8.85 & -8.78 & -2.98 & -9.18 & -8.86  \\ \hline
ANN & 6.46 & 0.00 & -2.32 & -2.40 & -2.32 & 3.48 & -2.72 & -2.41  \\ \hline
RBFN-RAN & 8.78 & 2.32 & 0.00 & -0.08 & 0.00 & 5.80 & -0.40 & -0.09  \\ \hline
RBFN-FCM & 8.85 & 2.40 & 0.08 & 0.00 & 0.08 & 5.87 & -0.33 & -0.01  \\ \hline
RBFN-KCM & 8.78 & 2.32 & 0.00 & -0.08 & 0.00 & 5.80 & -0.40 & -0.09  \\ \hline
BTE & 2.98 & -3.48 & -5.80 & -5.87 & -5.80 & 0.00 & -6.20 & -5.88  \\ \hline
MVE & 9.18 & 2.72 & 0.40 & 0.33 & 0.40 & 6.20 & 0.00 & 0.31  \\ \hline
NDTF & 8.86 & 2.41 & 0.09 & 0.01 & 0.09 & 5.88 & -0.31 & 0.00  \\ \hline

 \multicolumn{9}{|c|}{\textbf{p-value}}	\\ \hline
& \textbf{LOGR} &\textbf{ ANN} & \textbf{RBFN-RAN} & \textbf{RBFN-FCM} & \textbf{RBFN-KCM }& \textbf{BTE} & \textbf{MVE} & \textbf{NDTF } \\ \hline
LOGR & 1.00 & 0.00 & 0.00 & 0.00 & 0.00 & 0.00 & 0.00 & 0.00  \\ \hline
ANN & 0.00 & 1.00 & 0.00 & 0.00 & 0.00 & 0.24 & 0.00 & 0.00  \\ \hline
RBFN-RAN & 0.00 & 0.00 & 1.00 & 0.17 & 1.00 & 0.00 & 0.00 & 0.93  \\ \hline
RBFN-FCM & 0.00 & 0.00 & 0.17 & 1.00 & 0.17 & 0.00 & 0.26 & 0.12  \\ \hline
RBFN-KCM & 0.00 & 0.00 & 1.00 & 0.17 & 1.00 & 0.00 & 0.00 & 0.93  \\ \hline
BTE & 0.00 & 0.24 & 0.00 & 0.00 & 0.00 & 1.00 & 0.00 & 0.00  \\ \hline
MVE & 0.00 & 0.00 & 0.00 & 0.26 & 0.00 & 0.00 & 1.00 & 0.02  \\ \hline
NDTF & 0.00 & 0.00 & 0.93 & 0.12 & 0.93 & 0.00 & 0.02 & 1.00  \\ \hline

			\end{tabular}}
			
		}

		\subfloat[All metrics and Selected Metrics \label{testfs}]{
			\centering
			\renewcommand{\arraystretch}{1.1}
			\resizebox{8.5cm}{!}{				
				\begin{tabular}{|l|*{10}{c|}r}
					\hline
						\multicolumn{5}{|c|}{\textbf{Accuracy}} & \multicolumn{5}{|c|}{\textbf{F-Measure}} \\ \hline
					& \multicolumn{2}{|c|}{\textbf{Mean}}	& \multicolumn{2}{|c|}{\textbf{P-value}} & & \multicolumn{2}{|c|}{\textbf{Mean}}	& \multicolumn{2}{|c|}{\textbf{P-value}} \\ \hline					
					& \textbf{AM} & \textbf{SM }& \textbf{AM }& \textbf{SM} &  & \textbf{AM} & \textbf{SM }& \textbf{AM} & \textbf{SM} \\ \hline
				AM & 	0.00 & -3.15 & 1.00 & 0.00 & AM & 0.00 & -0.02 & 1.00 & 0.00 \\ \hline
				SM & 	3.15 & 0.00 & 0.00 & 1.00 & SM & 0.02 & 0.00 & 0.00 & 1.00 \\ \hline

				\end{tabular}}

			}

\end{table}	
\begin{figure}[h]
	\centering
	\subfloat[Classfication Techniques \label{necosttech}]{
		
		\includegraphics[width=8cm, height=4.0cm]{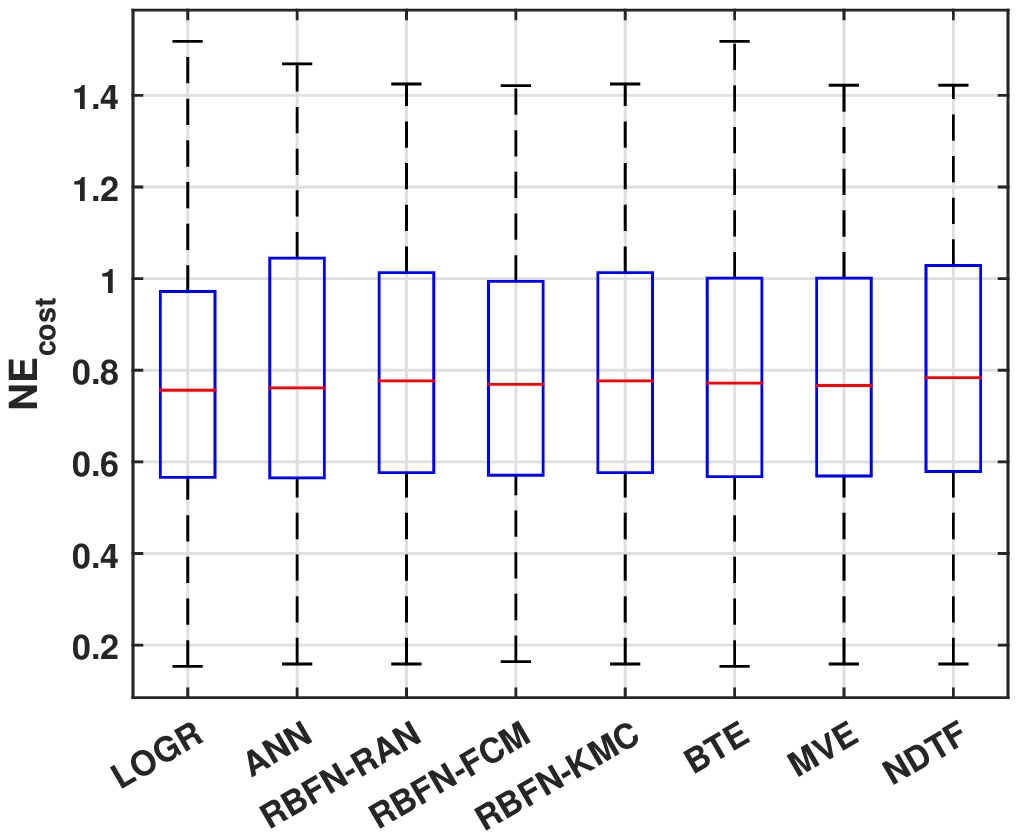} }
	
	\subfloat[Source code metrics\label{necostmetrics}]{
		
		\includegraphics[width=5cm, height=4.0cm]{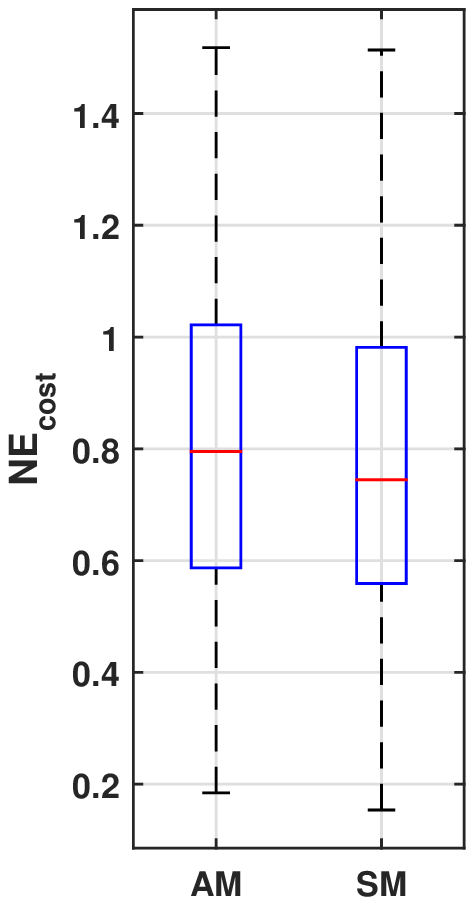} } 
	\caption{$ NE_{cost} $}
	\label{necostvalue}
\end{figure}
\begin{figure}[h!]
	\centering
	\includegraphics[width=8.5cm, height=3.8cm]{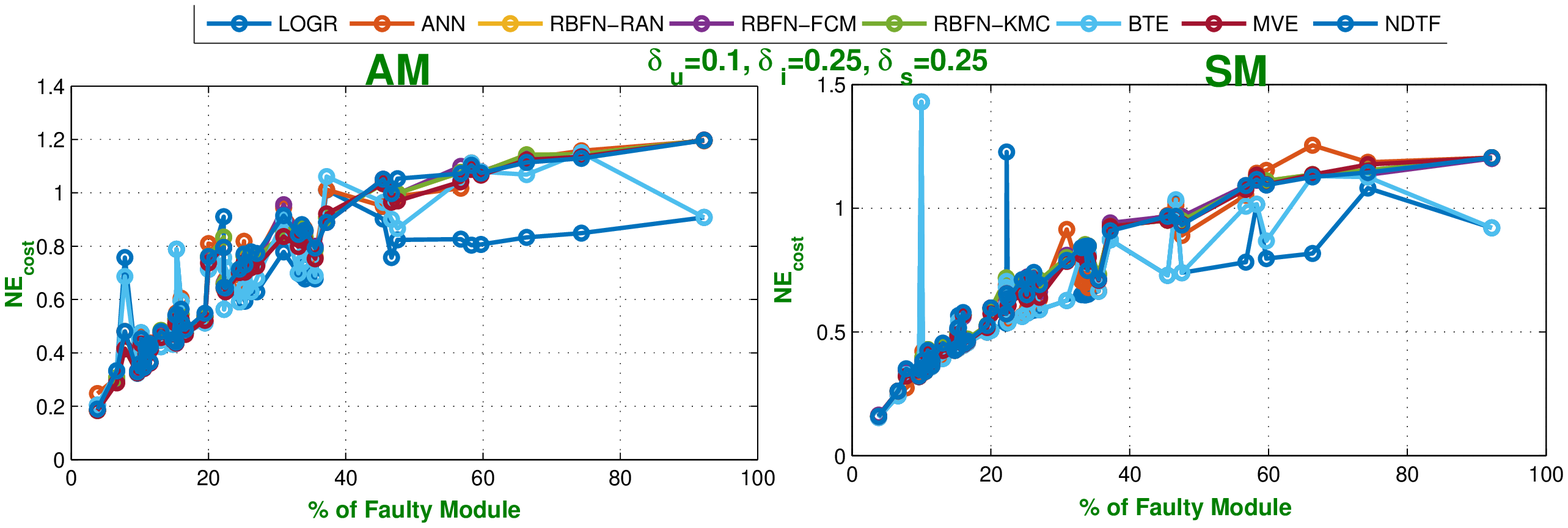}
	\caption{$NE_{cost}$ for $\delta_{u}=0.1, \delta_{i}=0.25, \delta_{s}=0.25$}
	\label{necostl}
\end{figure}
\begin{figure}[h!]
	\centering
	\includegraphics[width=8.5cm, height=3.8cm]{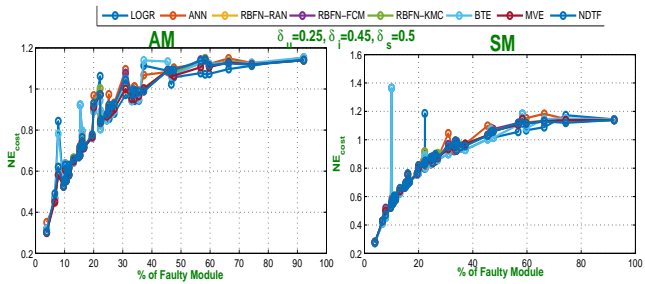}
	\caption{$NE_{cost}$ for $\delta_{u}=0.25, \delta_{i}=0.45, \delta_{s}=0.5$}
	\label{necostm}
\end{figure}
\begin{figure}[h!]	
	\centering
	\includegraphics[width=8.5cm, height=3.8cm]{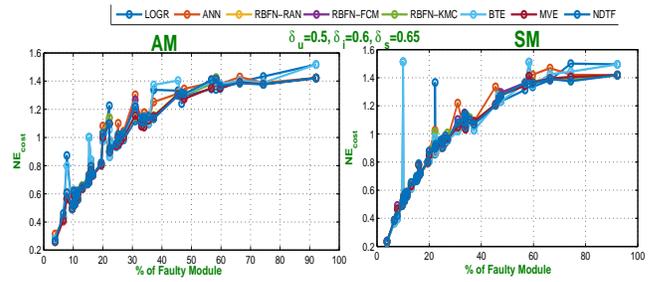}
	\caption{$NE_{cost}$ for $\delta_{u}=0.5, \delta_{i}=0.6, \delta_{s}=0.65$}
	\label{necosth}
\end{figure}
\begin{table}[h!]
	\caption{Threshold Value}
	\label{thrval}
	\centering
	\vspace{-1.0em}
	\subfloat[Training Methods\label{thrclass}]{
		\centering
		
		\begin{tabular}{|l|*{4}{c|}r}\hline			
			
			& \textbf{Const.}  & \textbf{Coeff.}  & \textbf{Threshold}   \\ \hline	\multicolumn{4}{|c|}{\boldmath$\delta_{u}=0.1, \delta_{i}=0.25, \delta_{s}=0.25$} \\ \hline
			
LOGR & -3.62 & 0.09 & 38.14  \\ \hline
ANN & -13.16 & 0.27 & 48.75  \\ \hline
RBFN-RAN & -23.98 & 0.48 & 49.75  \\ \hline
RBFN-FCM & -23.98 & 0.48 & 49.75  \\ \hline
RBFN-KCM & -23.98 & 0.48 & 49.75  \\ \hline
BTE & -4.71 & 0.08 & 61.48  \\ \hline
MVE & -23.98 & 0.48 & 49.75  \\ \hline
NDTF & -24.57 & 0.51 & 47.94  \\ \hline

			\multicolumn{4}{|c|}{\boldmath$\delta_{u}=0.25, \delta_{i}=0.45, \delta_{s}=0.5$} \\ \hline
LOGR & -6.67 & 0.17 & 39.64  \\ \hline
ANN & -11.98 & 0.33 & 36.18  \\ \hline
RBFN-RAN & -11.93 & 0.30 & 39.48  \\ \hline
RBFN-FCM & -16.92 & 0.42 & 39.91  \\ \hline
RBFN-KCM & -11.93 & 0.30 & 39.48  \\ \hline
BTE & -8.47 & 0.21 & 39.55  \\ \hline
MVE & -17.10 & 0.45 & 38.40  \\ \hline
NDTF & -16.92 & 0.42 & 39.91  \\ \hline

			\multicolumn{4}{|c|}{\boldmath$\delta_{u}=0.5, \delta_{i}=0.60, \delta_{s}=0.65$} \\ \hline
LOGR & -7.40 & 0.27 & 27.50  \\ \hline
ANN & -10.88 & 0.42 & 25.78  \\ \hline
RBFN-RAN & -11.56 & 0.46 & 25.17  \\ \hline
RBFN-FCM & -18.80 & 0.68 & 27.68  \\ \hline
RBFN-KCM & -11.56 & 0.46 & 25.17  \\ \hline
BTE & -7.98 & 0.28 & 28.00  \\ \hline
MVE & -12.13 & 0.44 & 27.47  \\ \hline
NDTF & -12.26 & 0.47 & 26.23  \\ \hline
		\end{tabular}
	}

	\subfloat[ All Metrics and Selected Metrics \label{thrsm}]{
		\centering
		
		\begin{tabular}{|l|*{4}{c|}r}\hline	
			& \textbf{Const.}  & \textbf{Coeff.}  & \textbf{Threshold}   \\ \hline		
			\multicolumn{4}{|c|}{\boldmath$\delta_{u}=0.1, \delta_{i}=0.25, \delta_{s}=0.25$} \\ \hline
			
			AM & -7.66 & 0.14 & 53.41  \\ \hline
			SM & -7.25 & 0.13 & 54.82  \\ \hline

			\multicolumn{4}{|c|}{\boldmath$\delta_{u}=0.25, \delta_{i}=0.45, \delta_{s}=0.5$} \\ \hline
		AM & -10.68 & 0.28 & 37.50  \\ \hline
		SM & -11.23 & 0.27 & 41.04  \\ \hline

			\multicolumn{4}{|c|}{\boldmath$\delta_{u}=0.5, \delta_{i}=0.60, \delta_{s}=0.65$} \\ \hline
		AM & -9.29 & 0.37 & 25.23  \\ \hline
		SM & -12.47 & 0.44 & 28.10  \\ \hline
\end{tabular}		
}
\end{table}
\subsection{Performance Evaluation Parameters}
We have considered two different set of metrics based on source code as input to design a model to estimate fault proneness of Java classes developed using 5 classification techniques i.e., LOGR, ANN, RBFN-RAN, RBFN-FCM, and RBFN-KMC and 3 ensemble methods i.e., BTE, MVE, and NDTF. Accuracy ($\%$) and F-Measure are considered as performance parameters to measure the performance of the developed fault prediction models.  Figure \ref{perclass} and Figure \ref{perclasssm} show the box-plot diagrams for each of the experimental results respectively enabling a visual comparison. The middle line of the boxes show the median of accuracy and F-Measure.
We apply 10-fold cross validation for all the combinations and the accuracy and f-measure metric values are summarized in the box blots. From the box-plot diagram, it can be inferred that:
\begin{itemize}
	\item In all cases, the selected set of source code metrics has a high median value. Based on the boxplots, SM produced the best result, i.e. the proposed software metrics validation method computes the best set of source code metrics for predicting faulty and non-faulty classes of object-oriented software as compared to all metrics.
	
	\item Among all classification, ensemble methods have outperformed as compared to individual models. Further, It is observed that MVE yields better results compared to other techniques.
\end{itemize}
\subsection{Comparison of results}
In this work, we have considered pairwise Wilcoxon signed rank test to determine which of the
classification techniques and selected sets of source code metrics work better or weather they all perform equally well. The use of Wilcoxon test without Bonferroni correction is not advisable because it does not take into account family-wise errors. In this work, we have considered Wilcoxon test with Bonferroni correction for comparison analysis.

\subsubsection{Classification Techniques}
Five different classification techniques and three different ensemble methods have been considered to develop a model to predict whether the class is faulty or not. Two different sets of metrics, one containing all metrics and one selected set using the proposed metrics validation techniques, have been considered as the input to develop fault prediction models over 45 different projects with two different performance parameters, i.e. accuracy, and F-Measure. In other words, for each technique a total number of two sets (one for each performance) is used, each with 90 data points ((1 feature selection method + 1 considering all features) * 45 datasets)). The results of the pair-wise comparisons of different training algorithms are shown in \ref{testcm}. 

Table \ref{testcm} contains two parts; the first part of shows the mean difference values and the second part shows the p-value between different pairs. The Bonferroni correction sets the significance cutoff at $\frac{\alpha}{n}$. In this study, eight different techniques have been considered for analysis, i.e. total number of twenty eight (28) different pairs are possible ($^{8 technique}C_{2}=8*7/2=28$) and all results are analyzed at a 0.05 significance level. Hence, we can only reject a null hypothesis if the p-value is less then $\frac{0.05}{28}=0.0018$. The null hypothesis of the Wilcoxon test is that there is no significant difference between the two techniques. From Table \ref{testcm}, it is evident that in most of the cases there is a significant difference between these approaches due to the fact that the p-value is smaller than 0.0018, out of 28 pairs of training methods, 19 are found to have significant results. From Table \ref{testcm}, it is also observed that the ensemble methods have outperformed when compared to individual models. Further, It is observed that MVE yields better results compared to other techniques.

\subsubsection{All metrics and Selected Metrics}
In this work, two different sets of metrics have been considered as the input to develop a model over 45 different object-oriented softwares. Eight different classification methods have been considered to develop a prediction model considering two different performance parameters, i.e. accuracy, and F-Measure. Consequently, for each set of metrics a total number of two sets (one for each performance measure) is used, each with 360 data points (8 classification techniques * 45 datasets). The results of Wilcoxon signed rank test analysis for performance parameters are summarized in Table \ref{testfs}. From Table \ref{testfs}, it may be observed that there is a significant difference between these approaches due to the fact that the p-value is smaller than 0.05. Yet, by judging the value of the mean difference, it is observed that the selected set of source code metrics using the proposed methods yields better results compared to all source code metrics.

\subsection{Cost analysis}	
In this experiment, the normalized fault removal cost approach suggested by Wagner \textit{et al.} \cite{a35} and the fault identification efficiencies for different testing phases suggested by Jones \cite{a20} have been used in designing of our cost evaluation model. Equations \ref{EqnEcost} and \ref{EqnTcost} are used to calculate the estimated fault removal cost ($E_{cost})$, and estimated testing cost ($T_{cost}$), respectively. 
Figure \ref{necostvalue} shows the normalized fault cost of different techniques and set of metrics. From Figure \ref{necosttech}, it can be seen that the MVE has low median value of $NE_{cost}$ as compare to other techniques. This shows that the fault prediction model developed using MVE method consume less fault removal cost as compare to other techniques. Similarly From Figure \ref{necostmetrics}, it is observed that model developed using selected set of metrics obtained less $NE_{cost}$ as compare to all metrics.  

Figures \ref{necostl} to \ref{necosth} depict the normalized fault removal cost ($NE_{cost}$) of fault prediction techniques for different values of $\delta_{u}$, $\delta_{i}$, and $\delta_{s}$. From these figures,  it is observed that as the percentage value of faulty classes increases, the fault-prediction technique tends to have a higher value of $ NE_{cost} $, i.e. fault prediction can be useful for the projects with percentage of faulty classes having less than certain threshold.
\subsection{Threshold Value}	
Once we identify the best performing model, We analyze the results to establish a relationship or extent of association between the $NE_{cost} $ and the percentage of faulty classes (FP). In this study, we use logistic regression approach as our dependent variable is dichotomous for the purpose of developing a statistical model to calculate the probability of usefulness of fault prediction techniques ($P_{fault}$). In logistic or logit regression, the dependent variable is binary and hence can take only two values. Therefore, we divide the dependent variable of a $ NE_{cost} $ into two groups: one group containing software for which fault prediction will be useful ($ NE_{cost}  < 1 $) and another group for which the fault prediction will not be useful ($ NE_{cost }  \geq 1 $). Table \ref{thrval} displays the constant, coefficient, and threshold values in terms of the percentage of faulty classes for different values of $\delta_{u}$ and $\delta_{s}$. We apply logit regression by setting a fixed threshold value of $0.5$. The threshold value implies that the fault prediction is useful if $P_{fault} < 0.5 $, otherwise the fault prediction is not useful. From Table \ref{thrval}, we can observe that the fault prediction is useful for the software projects having percentages of faulty classes less than a certain threshold. For example, Table \ref{thrval} reveals that the threshold value expressed in percentage for LOGR and ANN for the case of $\delta_{u}=0.1, \delta_{i}=0.25, \delta_{s}=0.25$ is $38.14$ and $48.75$ respectively.  
\section{Conclusions}
We conclude that the selected set of source code metrics has a high median value in terms of accuracy for all the classifier combinations in comparison to all metrics. This shows that identifying a subset of source code metrics is important. Our findings reveal that ensemble method learning algorithm outperforms individual classifiers. We observe that the MVE approach performs the best. We also observe that fault prediction model developed using MVE method consume less fault removal cost as compare to other techniques. We conclude that our fault prediction method is also effective for software projects with a percentage of faulty classes lower than the threshold value (low - 54.82\%, medium - 41.04\%, high - 28.10\%).
\balance
\bibliographystyle{abbrv}
\bibliography{sigproc}  

\begin{thebibliography}{10}

\bibitem{arisholm2010}
E.~Arisholm, L.~C. Briand, and E.~B. Johannessen.
\newblock A systematic and comprehensive investigation of methods to build and
  evaluate fault prediction models.
\newblock {\em Journal of Systems and Software}, 83(1):2--17, 2010.

\bibitem{bahler2000methods}
D.~Bahler and L.~Navarro.
\newblock Methods for combining heterogeneous sets of classifiers.
\newblock In {\em 17th Natl. Conf. on Artificial Intelligence (AAAI), Workshop
  on New Research Problems for Machine Learning}, 2000.

\bibitem{Basili}
V.~R. Basili, L.~C. Briand, and W.~L. Melo.
\newblock A validation of object-oriented design metrics as quality indicators.
\newblock {\em IEEE Transactions on Software Engineering}, 22(10):751--761,
  October 1996.

\bibitem{Batti}
R.~Battiti.
\newblock First and second-order methods for learning between steepest descent
  and newton's method.
\newblock {\em Neural Computation}, 4(2):141--166, 1992.

\bibitem{bian2007diversity}
S.~Bian and W.~Wang.
\newblock On diversity and accuracy of homogeneous and heterogeneous ensembles.
\newblock {\em International Journal of Hybrid Intelligent Systems},
  4(2):103--128, 2007.

\bibitem{a20}
J.~C.
\newblock Software quality in 2010: a survey of the state of the art.
\newblock In {\em Founder and Chief Scientist Emeritus}, 2010.

\bibitem{catal2009}
C.~Catal and B.~Diri.
\newblock A systematic review of software fault prediction studies.
\newblock {\em Expert systems with applications}, 36(4):7346--7354, 2009.

\bibitem{chen2009empirical}
J.-C. Chen and S.-J. Huang.
\newblock An empirical analysis of the impact of software development problem
  factors on software maintainability.
\newblock {\em Journal of Systems and Software}, 82(6):981--992, 2009.

\bibitem{chidamber1991}
S.~R. Chidamber and C.~F. Kemerer.
\newblock Towards a metrics suite for object oriented design.
\newblock {\em SIGPLAN Not.}, 26(11):197--211, Nov. 1991.

\bibitem{chidamber1991towards}
S.~R. Chidamber and C.~F. Kemerer.
\newblock Towards a metrics suite for object oriented design.
\newblock {\em SIGPLAN Not.}, 26(11):197--211, Nov. 1991.

\bibitem{doraisamy2008study}
S.~Doraisamy, S.~Golzari, N.~Mohd, M.~N. Sulaiman, and N.~I. Udzir.
\newblock A study on feature selection and classification techniques for
  automatic genre classification of traditional malay music.
\newblock In {\em ISMIR}, pages 331--336, 2008.

\bibitem{forman2003extensive}
G.~Forman.
\newblock An extensive empirical study of feature selection metrics for text
  classification.
\newblock {\em The Journal of machine learning research}, 3:1289--1305, 2003.

\bibitem{furlanello2003entropy}
C.~Furlanello, M.~Serafini, S.~Merler, and G.~Jurman.
\newblock Entropy-based gene ranking without selection bias for the predictive
  classification of microarray data.
\newblock {\em BMC bioinformatics}, 4(1):1, 2003.

\bibitem{hall2012}
T.~Hall, S.~Beecham, D.~Bowes, D.~Gray, and S.~Counsell.
\newblock A systematic literature review on fault prediction performance in
  software engineering.
\newblock {\em IEEE Transactions on Software Engineering}, 38(6):1276--1304,
  2012.

\bibitem{a36}
R.~Huitt and N.~Wilde.
\newblock Maintenance support for object-oriented programs.
\newblock {\em IEEE Transactions on Software Engineering}, 18(12):1038--1044,
  1992.

\bibitem{kumar2016quasoc}
L.~Kumar, S.~K. Rath, and A.~Sureka.
\newblock Predicting quality of service (qos) parameters using extreme learning
  machines with various kernel methods.
\newblock In {\em 4th International Workshop on Quantitative Approaches to
  Software Quality}, page~11, 2016.

\bibitem{kumar2017isec}
L.~Kumar, S.~K. Rath, and A.~Sureka.
\newblock Empirical analysis on effectiveness of source code metrics for
  predicting change-proneness.
\newblock In {\em Proceedings of the 10th Innovations in Software Engineering
  Conference}, pages 4--14. ACM, 2017.

\bibitem{kumar2017malt}
L.~Kumar, S.~K. Rath, and A.~Sureka.
\newblock Using source code metrics to predict change-prone web services: A
  case-study on ebay services.
\newblock In {\em Machine Learning Techniques for Software Quality Evaluation
  (MaLTeSQuE), IEEE Workshop on}, pages 1--7. IEEE, 2017.

\bibitem{compsac2017}
L.~Kumar, R.~Santanu, and A.~Sureka.
\newblock An empirical analysis on effective fault prediction model developed
  using ensemble methods.
\newblock {\em 2017 IEEE 41st Annual Computer Software and Applications
  Conference (COMPSAC)}, 2017.

\bibitem{henry}
W.~Li and S.~Henry.
\newblock Maintenance metrics for the {Object-Oriented} paradigm.
\newblock In {\em Proceedings of First International Software Metrics
  Symposium}, pages 52--60, 1993.

\bibitem{a35}
W.~S.
\newblock A literature survey of the quality economics of defect-detection
  techniques.
\newblock In {\em Proceedings of the ACM/IEEE International Symposium on
  Empirical Software Engineering (ISESE)}, pages 194--203, 2006.

\bibitem{xu2015efspredictor}
B.~Xu, D.~Lo, X.~Xia, A.~Sureka, and S.~Li.
\newblock Efspredictor: Predicting configuration bugs with ensemble feature
  selection.
\newblock In {\em Software Engineering Conference (APSEC), 2015 Asia-Pacific},
  pages 206--213. IEEE, 2015.

\bibitem{zhang2016direct}
C.~Zhang, H.~Wei, L.~Xie, Y.~Shen, and K.~Zhang.
\newblock Direct interval forecasting of wind speed using radial basis function
  neural networks in a multi-objective optimization framework.
\newblock {\em Neurocomputing}, 2016.

\end{thebibliography}

\end{document}